# Two-Step Tapering-Collapse Method Enables Element-Interdiffused Cladding for Enhanced Laser Amplification in Yb:YAG Single Crystal Fibers

*Xiangfei Zhu[1,2,†], Xiaofei Ma[1,2,†,\*], Chenxin Gao[1,2,\*], Tao Wang[3], Zhen Huang[1,2], Jiajia Zeng[1,2], Dongran Shi[1,2], Xuanfeng Zhou[1,2], Baolai Yang[1,2], Zilun Chen[1,2], Hu Xiao[1,2], Feng Xiong[1,2], Pengfei Ma[1,2], Jian Zhang[3], Zhitai Jia[3], Zefeng Wang[1,2,\*], Xutang Tao[3]*

[1]College of Advanced Interdisciplinary Studies, National University of Defense Technology, Changsha 410073, China
[2]Nanhu Laser Laboratory, National University of Defense Technology, Changsha 410073, China
[3]State Key Lab of Crystal Materials, Shandong University, Jinan 250100, China
[†]These authors contributed equally to this work.
[\*]Corresponding authors: xiaofeima_@nudt.edu.cn, chenxingao24@nudt.edu.cn, zefengwang_nudt@163.com

**Abstract:** The development of high-power single crystal fiber (SCF) lasers is critically hindered by the lack of a reliable cladding scheme to confine the optical mode and ensure beam quality. Here, we propose and demonstrate a two-step tapering-collapse method for the first time to fabricate a high-quality cladding on Yb:YAG SCFs based on elemental interdiffusion. This in-situ formed crystalline transition layer with a graded refractive index effectively suppresses lattice mismatch and abruptly mitigates core-cladding interfacial stress. Consequently, the numerical aperture of the SCF is significantly reduced from 0.280 to 0.199. In a master oscillator power amplifier configuration, the clad SCF delivers a remarkable 46.7% enhancement in slope efficiency compared to its bare counterpart, accompanied by a substantially improved near-field beam profile. This work establishes a facile and effective route to high-performance clad SCFs, unlocking their full potential for next-generation extreme-condition lasers.

## 1. Introduction

As emerging low-dimensional functional crystalline materials, single crystal fibers (SCFs) have found extensive applications in harsh environment sensing[1,2], radiation detection[3,4], medical devices[5], and high-power laser systems[6–8]. Among these fields, high-power laser technology is evolving rapidly with escalating demands for higher output power, better beam quality, and more stable operation in scientific research, industrial manufacturing, and national defense fields[9–14]. However, conventional laser gain media face inherent limitations[11]. Silica fibers suffer from low thermal conductivity and high stimulated Brillouin scattering gain, which restrict power scaling to the kilowatt level[15,16]. Bulk crystals, despite their excellent material properties, exhibit inefficient heat dissipation due to their low surface-to-volume ratio, leading to thermal lensing and beam quality degradation under high pump power[17]. SCFs effectively overcome these drawbacks by combining crystalline material properties with a waveguide structure, positioning them as key candidates for next-generation high-power laser gain media[18–20].

Nevertheless, a long-standing bottleneck persists in the entire SCFs field. Unlike silica fibers that can be readily cladded via mature "jacketing-drawing" processes, SCFs possess inherently high and fixed melting points[21], precluding direct adoption of such flexible cladding strategies, leading to the lack of effective optical confinement. This not only induces severe degradation of output beam quality but also limits compatibility with low numerical aperture (NA) fiber systems, thereby restricting further improvements in laser power and brightness[22].

Accordingly, introducing a low-refractive-index cladding structure into SCFs represents a key technical strategy to optimize their waveguide characteristics and enhance beam quality. Existing cladding preparation schemes are primarily categorized into two classes[18]. One is modified cladding, which constructs a core-cladding waveguide structure by modulating the refractive index distribution of the fiber surface, such as introducing nanoscale cavities in the surface of SCF via hydrogen ion implantation[23] or femtosecond laser inscription of microstructures[24]. The other is additional cladding, which involves depositing or compounding external materials on the core surface. Typical methods include glass cladding via co-drawing laser-heated pedestal growth (CDLHPG)[25,26], polymer cladding via light-curing[27], metal cladding via thermal curing[28], and single-crystal/polycrystalline cladding via hydrothermal growth[29], liquid-phase epitaxy[30] or physical vapor deposition[31].

Each of these approaches, however, suffers from inherent drawbacks. Modified claddings formed by refractive index modulation such as laser micromachining, maintain a homogeneous crystalline phase with the core, enabling maximum utilization of the intrinsic material advantages of SCFs, however, damaging the crystal structure and introduces non-negligible scattering losses. The fabrication of glass, polymer, and metal claddings is relatively straightforward, but due to lattice mismatch and thermal expansion coefficient mismatch between the cladding and core, such claddings are prone to delamination from the core under high-power laser operating conditions, failing to meet the requirements of high-reliability applications. To address the core-cladding interfacial mismatch, the single crystal derived fiber is developed by CDLHPG method, enabling splicing with silica fiber to form an all-fiber structure[32,33], nevertheless, fundamentally departing from the essence of an SCF and sacrificing the superior material properties of the crystal.

Notably, during the fabrication of glass claddings on $Cr^{4+}$ doped yttrium aluminum garnet via CDLHPG with the assistance of sapphire tube, researchers observed that due to solid-phase diffusion, the fused glass capillary softened, shrank, and formed a mixed inner cladding with YAG which reduced the transmission loss from 0.6 dB/cm to 0.02 dB/cm[34]. Although the process was complex, it demonstrated the potential application value of such hybrid structures in laser amplification. Recently, researchers have successfully fabricated 3 cm-long single crystal fibers via the CDLHPG method by cladding high-refractive-index flint glass on the surface of YAG crystals, which theoretically enables single-mode transmission in the 1.5 to 3 μm band[35]. However, this method suffers from the inaccessibility of raw materials and, essentially, fails to overcome the lattice mismatch between the fiber core and cladding.

In this work, we propose a two-step tapering-collapse method to fabricate an element-interdiffused cladding on a Yb:YAG SCF with a diameter of 400 μm for the first time, followed by comprehensive morphological and phase and NA characterizations as well as laser amplification performance verification. The resulting cladding exhibits a crystalline structure with graded elemental distribution, effectively suppressing lattice mismatch between core and cladding, meanwhile, reducing the numerical aperture from 0.280 to 0.199. Furthermore, we construct a master oscillator power amplifier (MOPA) laser system based on clad single crystal fiber (CSCF), achieving a slope efficiency improvement of over 46.7% compared to bare single crystal fiber (BSCF), accompanied by remarkably improved beam quality. This approach enables the fabrication of CSCFs with a crystalline core and an element-interdiffused crystalline inner cladding and resolves the critical challenges of poor thermal stress matching at the core-cladding interface and inferior mechanical properties of the cladding, exhibiting great applicability for laser applications under extreme conditions including high-power laser systems, aerospace laser communication, and extreme environment laser processing.

## 2. Results and Discussion

### 2.1 Characterization of the CSCF

*2.1.1. Morphology and Phase Characterization of the CSCF*
The CSCF is fabricated via the two-step tapering-collapse method, as illustrated in **Figure 1**, with the detailed fabrication procedure provided in the Experimental Section. To characterize the morphology, phase composition, elemental distribution, and core-cladding interfacial quality of the CSCF, we performed optical microscopy imaging, scanning electron microscopy (SEM) imaging, Raman spectroscopy, energy dispersive spectroscopy (EDS), and electron probe microanalysis (EPMA) on the CSCF. The optical microscopic image of the end face of CSCF reveals a distinct three-layer structure (from the inside out: core, inner cladding, and outer cladding with diameters of 356.9, 377.6 and 534.5 μm, respectively), as displayed in **Figure 2**a. To further elucidate the interlayer bonding quality and validate the endface polishing quality of the CSCF, SEM characterizations were conducted. The results reveal an excellent interlayer bonding quality and a nanoscopically flat CSCF endface, which renders the fiber well-suited for laser amplification applications. The diameter scanning results demonstrate that the CSCF has an average diameter of 535.1 μm and fiber length of 6.3 cm with a diameter fluctuation of only 1.554%, confirming the excellent cladding uniformity achieved by the two-step tapering-collapse method.

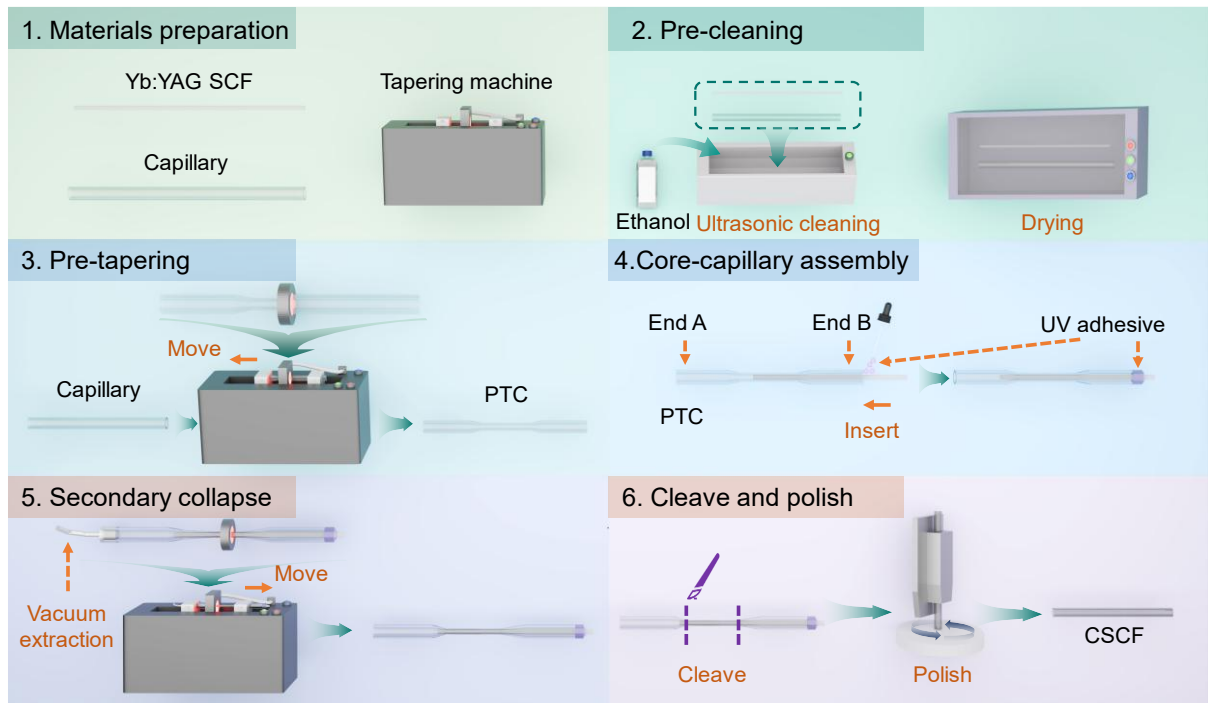

**Figure 1.** Schematic illustration of the two-step tapering-collapse method for cladding fabrication on SCFs. PTC: pre-tapered capillary. CSCF: clad single crystal fiber. The fabrication process involves six sequential steps. Step 1: preparation of Yb:YAG SCF, fluorine-doped glass capillary and fiber tapering machine as raw materials and equipment. Step 2: pre-cleaning of the aforementioned materials. Step 3: pre-tapering of the capillary to form a waist region matching the SCF diameter. Step 4: coaxial assembly of SCF and PTC with UV adhesive for airtight sealing. Step 5: high-temperature secondary collapse bonding of the PTC under a low pressure environment to form the CSCF. Step 6: precision cleaving and polishing of the as-prepared CSCF to achieve device-level fabrication.

To characterize the crystalline state and material composition of the CSCF across radial positions, five different locations on the end face of CSCF were sampled, which were located at the center of core (location 1), edge of core (location 2), inner cladding (location 3), inner side of the outer cladding (location 4) and edge of the outer cladding (location 5), as illustrated in Figure 2b. Then comparative Raman spectroscopy measurements were performed on the Yb:YAG BSCF, five sampling locations at the end face of CSCF, and fluorine-doped glass capillary, with the results illustrated in Figure 2c. The Raman spectrum of location 1 and 2 are in perfect agreement with those of the Yb:YAG BSCF, with characteristic peaks at 160 cm$^{-1}$, 216 cm$^{-1}$, 257 cm$^{-1}$, 367 cm$^{-1}$, 400 cm$^{-1}$, 541 cm$^{-1}$, 554 cm$^{-1}$, 688 cm$^{-1}$ and 713 cm$^{-1}$, corresponding to the typical vibration modes of YAG crystals, with no significant peak broadening, confirming the excellent crystallinity of core during the collapse process. The Raman spectrum of location 3 shows reduced intensity of YAG peaks with inverted relative intensities of the 367 cm$^{-1}$ and 400 cm$^{-1}$ peaks, attributed to lattice distortion from elemental diffusion. The Raman spectrum of location 4 and 5 exhibit an amorphous spectrum similar to that of the glass capillary, with weak YAG peaks observed at location 4, indicating diffusion-induced elemental penetration into the outer cladding.

The SEM and EDS were employed to analyze interfacial bonding and elemental distribution. The local SEM image shown in Figure 2d confirms the reliability of the core-cladding integration. The EDS mapping displayed in Figure 2e-g shows that Y element is concentrated in the core, gradually decreasing in the inner cladding, and absent in the outer cladding, while Si element is abundant in the outer cladding, weakly present in the inner cladding, and undetectable in the core, directly confirming that the inner cladding forms through interdiffusion between YAG and fluorine-doped silica.

To further quantitatively analyze the elemental content at different radial positions of the CSCF, an EPMA was employed for characterization, with the results displayed in Figure 2h. The core region is dominated by Yb, Y, Al, and O elements with stable contents. In the inner cladding region (location 3), the contents of Y and Al elements decrease significantly while Si and F elements begin to appear, presenting a gradient distribution of elements. The outer cladding region (location 4 and 5) is dominated by Si, O, and F elements with only trace amounts of residual Y and Al elements. These quantitative results are consistent with the observation of weak YAG peaks in the Raman spectrum, further confirming the occurrence of elemental interdiffusion and the formation of a graded transition layer.

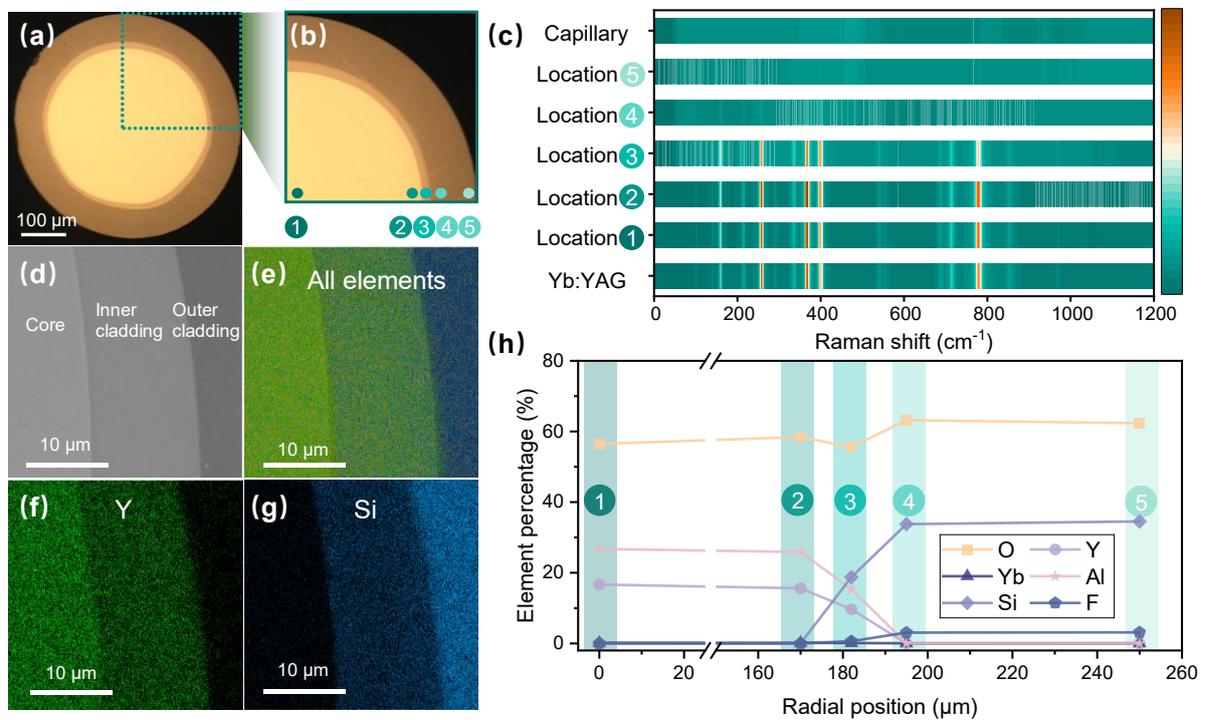

**Figure 2.** Morphology and phase characterization of CSCF. (a) Optical microscopic image of the end face of CSCF. (b) Magnified optical microscope image of the region within the green box in (a) , showing five sampling locations at the center of core, the edge of core, the inner cladding, the inner side of the outer cladding and the edge of the outer cladding, respectively. (c) Raman spectrum of the Yb:YAG BSCF, five sampling locations of the end face of CSCF and the fluorine-doped glass capillary. (d) Local SEM image of the CSCF end face. (e) Elemental distribution mapping of all elements. (f) Y element distribution mapping of the CSCF local end face. (g) Si element distribution mapping of the CSCF local end face. (h) Elemental contents at five different locations shown in (b) of the CSCF end face.

*2.1.2. Numerical aperture characterization of the CSCF*

To evaluate the influence of the aforementioned element-interdiffusion layer on the NA of the SCF, a NA measurement system was constructed, as illustrated in **Figure 3**a, with the detailed description of the experimental setup in the Experimental Section. Firstly, to validate the accuracy of the NA measurement setup, two different commercial single-clad silica fibers with known NA of 0.15 and 0.22 were employed as FUTs for test. The results, presented in Figure 3b, indicate that the measured NAs of the two fibers are 0.202 and 0.252 respectively, exhibiting a strong linear correlation with their nomimal NAs and thus confirming the reliability of the measurement system. A linear calibration was then performed between the nominal and measured NAs of the two fibers to establish a calibration function.

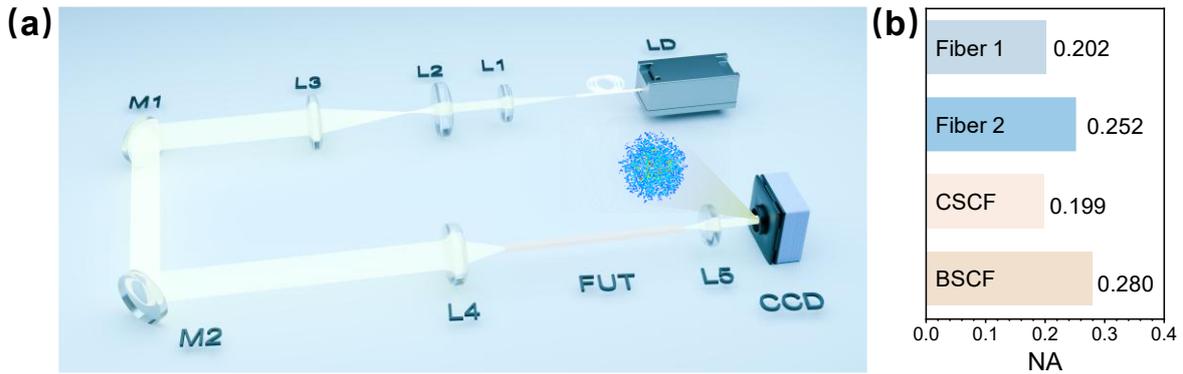

**Figure 3.** NA characterization of the CSCF and BSCF. LD: laser diode. L1-5: lens1-5. M1, M2: mirror1, mirror2. FUT: fiber under test. CCD: charge coupled device. (a) Schematic diagram of experimental NA measurement system. (b) NA measurement results of different fibers, including the fiber 1 and fiber 2 of single-clad silica fiber with known NA of 0.15 and 0.22 respectively, the CSCF and the BSCF.

The CSCF and BSCF were then tested using the same procedure to obtain their speckle diameters, and their NAs are calculated through the equation $NA = D / 2f_5$. Applying the calibration function, the actual NA values of CSCF and BSCF are determined to be 0.199 and 0.280, respectively, confirming that the NA of SCFs can be significantly reduced by fabricating the cladding structure. According to the theoretical principles governing the guided modes of multimode fibers, a reduction in NA leads to a decrease in the number of guided modes supported by the fiber. Consequently, when the CSCF is employed as laser gain media, the fiber yields improved output laser beam quality.

## 2.2. Laser Amplification Performance of the CSCF

To evaluate the impact of cladding fabrication on the laser gain performance of SCFs, a MOPA laser system was constructed, with the experimental configuration schematically illustrated in **Figure 4** and details provided in the Experimental Section.

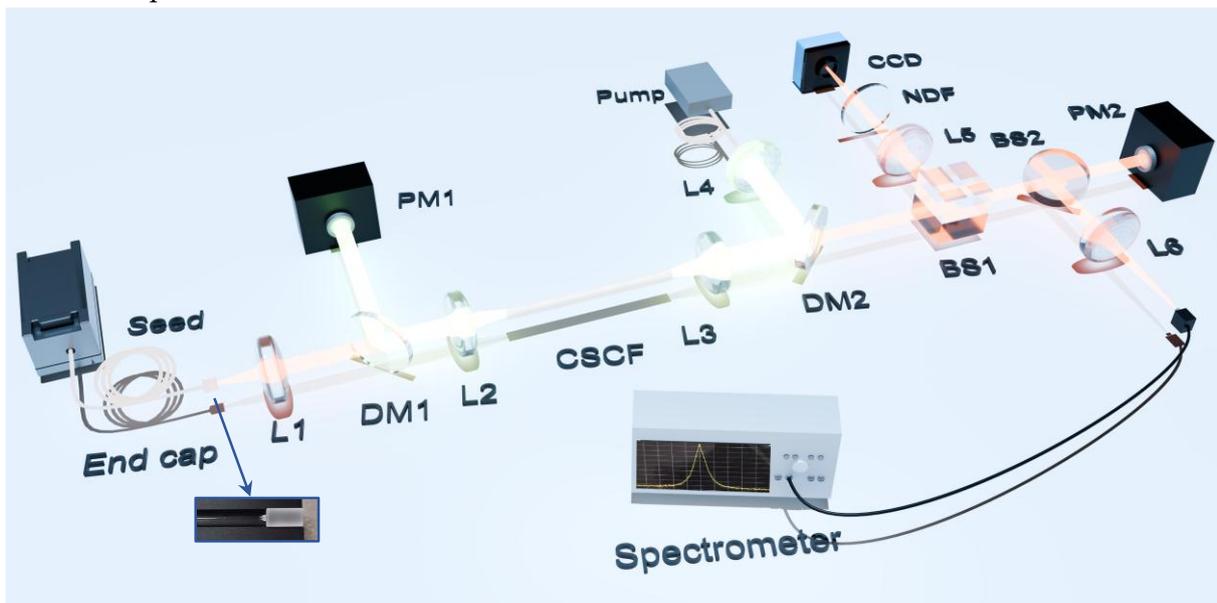

**Figure 4.** Schematic diagram of MOPA laser system (Inset: photograph of the end cap spliced with the seed source). L: lens. DM: dichroic mirror. PM: power meter. BS: beam splitter. CSCF: clad single crystal fiber. NDF: neutral density filter. CCD: charge coupled device. The gain medium of the amplifier is the CSCF fabricated as described above. The pump laser is coupled into the CSCF via lenses L4 and L3 in the form of backward end-pumping, while the seed laser is coupled into the CSCF through lenses L1 and L2 to realize power amplification. The amplified signal laser is detected and characterized by subsequent equipment after passing through the dichroic mirror DM2.

Based on the aforementioned amplifier structure, laser amplification experiments were conducted on the CSCF and BSCF respectively, with the experimental results presented in **Figure 5**. Figure 5a and c depict the power amplification characteristic of the CSCF and BSCF respectively with the power of pump laser ranging from 4 W to 8W. The results demonstrate that the output power of the CSCF is consistently higher than that of the BSCF at the same pump power across different seed laser power, verifying that the energy extraction efficiency of the CSCF is significantly superior to that of the BSCF. With the increase in seed laser power, the slope efficiencies of both fibers increase and reach their maximum at a seed laser power of 7 W. At their respective maximum efficiency operating points (seed laser power of 7 W), the slope efficiency of the CSCF (14.42%) is improved by 46.7% compared with that of the BSCF (9.48%), directly verifying the boosting effect of the cladding structure on amplification performance.

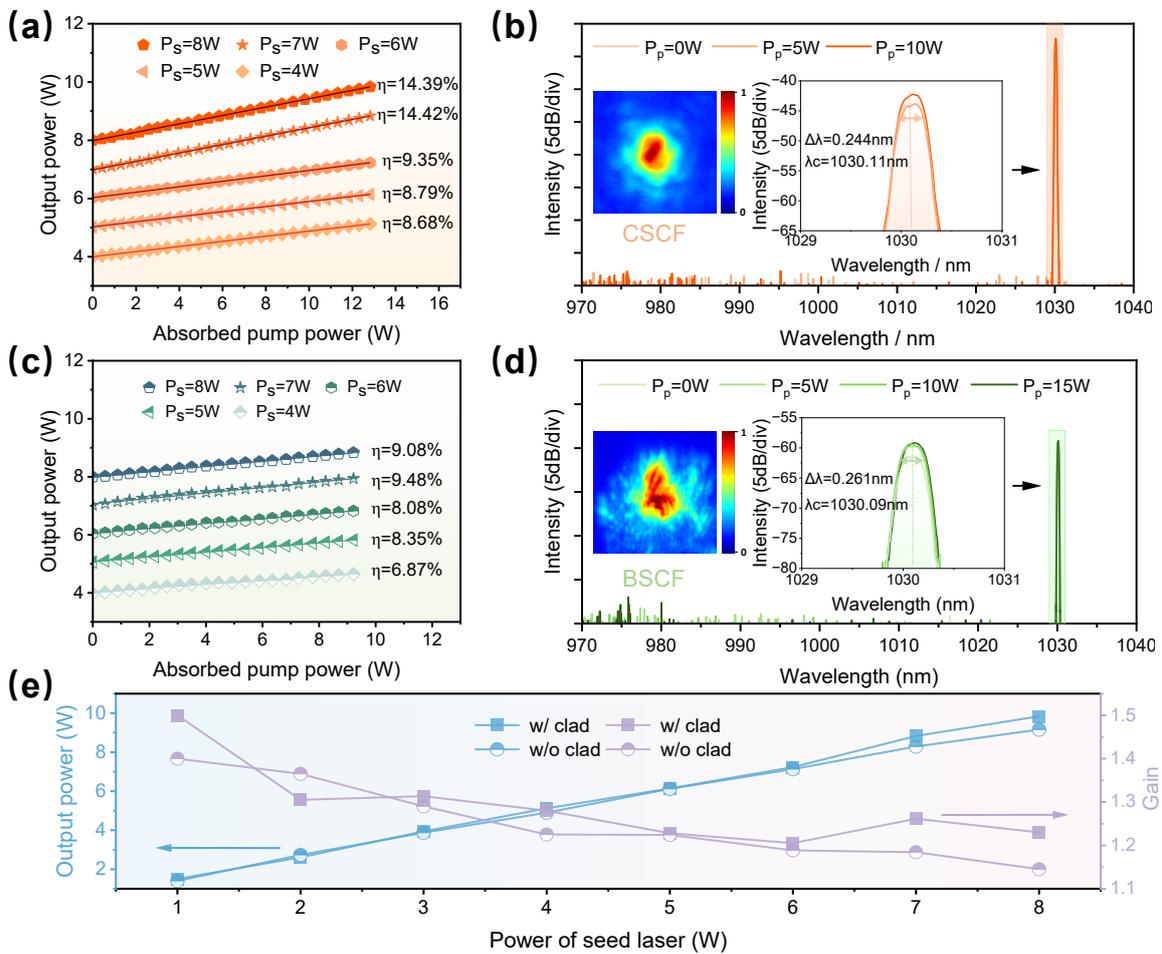

**Figure 5.** Laser amplification performance of the CSCF and BSCF. (a) Output power as a function of absorbed pump power for the CSCF at seed power ranging from 4 W to 8 W. (b) Output spectra of the amplified signal laser under the 8W seed laser with incident pump power of 0 W, 5 W and 10 W (Inset: Near-field spot profile of the amplified signal laser at an output power of 9.83 W). (c) Output power as a function of absorbed pump power for the BSCF at seed power ranging from 4 W to 8 W. (d) Output spectra of the amplified signal laser under the 8 W seed laser with incident pump power of 0 W, 5 W, 10 W and 15 W (Inset: near-field spot profile of the amplified signal laser at an output power of 8.83 W). (e) Comparisons of maximum output power and gain between the CSCF and BSCF at different power of seed laser.

Figure 5b and d illustrate the output spectral characteristics of the two laser amplifiers, with the insets showing the near-field beam profiles at the maximum output power. Within the watt-level output power range, the output spectrum of both fiber amplifiers exhibit no significant shift in the central wavelength and no broadening of the linewidth, indicating that no prominent nonlinear optical effects are excited during the amplification process. From the analysis of beam profile characteristics, the output beam of the BSCF suffers from obvious distortion, which can be attributed to parasitic mode oscillation induced by residual internal stresses of the material and fiber surface scattering. In contrast, the output beam of the CSCF displays excellent symmetry with remarkably improved beam quality, confirming that the proposed cladding fabrication scheme enables efficient mode field regulation, effectively suppresses high-order modes, and optimizes the output mode.

As shown in Figure 5e, we systematically investigated the maximum output power and gain of the CSCF and BSCF under maximum pump power by varying the seed laser injection power. Owing to the short fiber length, the pump light absorption of the fiber is insufficient, resulting in a relatively low slope efficiency of the amplifier. Notably, the maximum output power and gain of the CSCF are significantly higher than those of the BSCF under, directly validating the role of the proposed cladding fabrication scheme in enhancing the laser characteristics of BSCF.

## 3. Conclusion

In this study, a two-step tapering-collapse method for SCF cladding fabrication is proposed for the first time, and element-interdiffused claddings are successfully prepared on Yb:YAG SCFs. Morphology and phase characterization results demonstrate that an elemental diffusion layer is formed between the YAG crystal core and silica of the fiber under a high-temperature thermal field. Both the core and inner cladding exhibit a crystalline state, preserving the excellent physicochemical properties of YAG crystals, while effectively mitigating lattice mismatch and abrupt refractive index change, thereby enhancing the anti-interference capability of the SCF. Following cladding fabrication, the NA of the fiber is reduced from 0.280 to 0.199, confirming the cladding's functions in optical field confinement and high-order mode suppression.

The slope efficiency of the CSCF-based laser amplifier is improved by 46.7% compared with that of the BSCF-based amplifier counterpart with significant improvement in the output near-field beam profile. This elemental diffusion-based cladding fabrication strategy addresses the lattice mismatch between core and cladding, a key challenge in fabricating glass claddings for SCFs, and the sacrifice of intrinsic single crystal properties of the core in single crystal derived fibers. Furthermore, the thickness and refractive index distribution of the inner cladding can be tailored by further optimizing parameters such as the fluorine doping ratio of the capillary, the diameter of the pre-tapered region, the collapse rate, and the thermal field temperature and so on. Notably, the two-step tapering-collapse method developed in this work boasts the merits of low cost, facile operation and customizable parameters, and demonstrates great applicability for extreme-condition laser applications such as high-power laser amplification systems, aerospace laser communication and industrial extreme-environment laser processing.

## 4. Experimental Section

*Cladding Fabrication for Yb:YAG SCF*: The operational procedure of the two-step tapering-collapse method is illustrated in Figure 1, comprising six sequential processes: materials preparation, pre-cleaning, pre-tapering, core-capillary assembly, secondary collapse, cleaving and polishing. Firstly, materials and equipment were prepared, including a 1 at.% Yb:YAG SCF with a diameter of 400 μm grown through the laser-heated pedestal growth (LHPG) method, a commercial fluorine-doped glass capillary (Heraeus, TNU*) with inner/outer diameters of 800/1100 μm and a commercial fiber tapering machine (JUHERE, MT-180), as shown in step 1 of Figure 1. To ensure core-cladding bonding quality and reduce interfacial optical loss, the SCF and capillary were ultrasonically cleaned in 99% anhydrous ethanol for 15 min, followed by low-temperature drying to eliminate contaminants and residual solvent before the fabrication. After the pre-cleaning process, the capillary was horizontally passed through the graphite electrodes of the tapering machine, with both ends secured on precision translation stages. A uniform thermal field generated by the electrodes was subsequently applied to soften the capillary, and axial tapering was achieved via programmed precise velocity differences between the two translation stages, yielding a well-defined "tapered region–waist region–tapered region" structure, as illustrated in step 3 of Figure 1. The inner diameter of the waist region of the pre-tapered capillary (PTC) was controlled at approximately 410 μm (slightly larger than the diameter of SCF), providing sufficient assembly margin for subsequent core-cladding integration. Simultaneously, axial tension during tapering was monitored by a tension sensor and regulated through negative feedback to ensure diameter uniformity of the PTC waist regions. Subsequently, the SCF was axially inserted from the End-B of the PTC and guided through the waist region, after which the End-B of the capillary was sealed with UV adhesive to form a core-capillary assembly and maintain coaxial fixation of the SCF and capillary, as diagramed in step 4 of Figure 1. In the fifth step, the assembly was fixed on translation stages, with End-A connected to a vacuum pump for 5 minutes of pre-evacuation to create a negative pressure environment and further facilitate tight bonding between the PTC and SCF while suppressing bubble defects at the core-cladding interface. Under the high temperature environment generated by the graphite electrodes, the SCF and fluorine-doped glass capillary were tightly bonded at a collapse rate of 0.3 mm/s, forming the CSCF with continuous vacuum evacuation maintained throughout the process. Finally, for optical characterization and mitigation of scattering losses arising from end face irregularities, cleaving and end face polishing were performed to optimize the end face quality of the fabricated CSCF as shown in step 6 of Figure 1. Both ends werer cleaved with a diamond scribe to obtain initial flat surfaces, followed by end face polishing using a precision polishing system (NEOFIBO, NEOPL-1800A).

*Morphology and Phase Characterizations of the CSCF*: The diameter fluctuation of the CSCF is measured by the diameter scanning system of a commercial fiber tapering machine (3SAE Technologies, CMS). The SEM (TESCAN, MIRA4 LMH) is used to further clarify the interlayer bonding quality and verify the end face polishing quality. Then comparative Raman spectroscopy measurements are performed using a Raman spectrometer (RENISHAW, inVia Raman Microscope). The SEM and EDS (Oxford Instruments, One Max 50) are employed to analyze interfacial bonding and elemental distribution. To further quantitatively analyze the elemental content at different radial positions of the CSCF, an EPMA (SHIMADZU, EPMA-1720T) is employed for characterization.

*NA Measurements*: The schematic diagram of the NA measurement system is shown in Figure 3a. The laser source is a 976 nm laser diode with a core diameter of 105 μm and NA of 0.22, whose multimode property facilitates high-order mode excitation in the fiber under test (FUT). The test laser is collimated by lens L1 (f=20 mm) and then expanded via the beam expander system consisting of lenses L2 (f=30 mm) and L3 (f=60 mm). The expanded laser is redirected by two reflectors M1 and M2, and subsequently coupled into the FUT through lens L4 (f=15 mm). After exiting from the FUT, the output beam diverges rapidly and is then collimated by lens L5 (f=15 mm) before being detected by a charge coupled device (CCD) camera (ToupTek, IUA4100KPA). The two commercial fibers used for measurements are Fiber 1: Nufern MM-S200/220-15A and Fiber 2: Nufern MM-S200/220-22A.

To avoid underestimating NA due to insufficient excitation of high-order modes, the position of the incident spot was adjusted to scan 25 positions (5×5 grids) on the incident end face to excite high-order modes. The 25 output speckle patterns were superimposed to obtain the final output speckle. Using 5% of the maximum intensity as the threshold, the minimum circumscribed circle diameter of the speckle was calculated, and NA was determined by: $NA = D/2f_5$, where $f_5$ is the focal length of lens L5, and $D$ is the diameter of speckle.

*Laser Amplification*: To investigate the influence of the element-interdiffused cladding on the laser performance of SCFs, a laser amplifier setup was constructed, as illustrated in Figure 4. A home-made 1030 nm narrow-linewidth all-fiber laser oscillator serves as the seed source, with a core diameter of 15 μm and an NA of 0.080. To mitigate amplified spontaneous emission (ASE) reflection-induced damage to the seed source, the output fiber is spliced with a home-made end cap ( the insert in Figure 4). The pump source is a 976 nm commercial fiber-coupled laser diode with a core diameter of 105 μm and an NA of 0.220. To improve the amplification efficiency of the signal laser, the propagation processes of the signal and pump lasers within the fiber are simulated via the ray tracing method, and the focal lengths of the coupling lenses L1, L2, L3 and L4 are optimized. The simulation results indicate that to achieve the maximum spatial overlap between the two laser beams the spot sizes of the signal laser and pump laser focused on the CSCF end face should be identical. The pump laser is coupled into the CSCF via lenses L4 (f=60 mm) and L3 (f=75 mm) in the form of backward end-pumping. Compared with forward pumping, this configuration exhibits a lower pump laser utilization efficiency, while effectively reducing the thermal load on the fiber end face and the risk of end face damage. The seed laser is collimated by lens L1 (f=30 mm) after exiting the fiber endcap and then transmitted through a long-pass dichroic mirror (DM1) and focused into the CSCF by lens L2 (f=300 mm) for power amplification. The amplified signal laser is collimated by lens L3 (f=75 mm) after exiting the CSCF and outputs through dichroic mirror DM2. 1% of the signal laser is reflected by beam splitter BS1 (R:T=1:99) for near-field beam profile characterization and another 1% is reflected by BS2 (R:T=1:99) for spectral measurement. The power of the transmitted laser is measured by power meter.

## Conflict of Interest
The authors declare no conflict of interest.

## Data Availability Statement
The data that support the findings of this study are available from the corresponding author upon reasonable request.